\documentclass[conference]{IEEEtran}
\IEEEoverridecommandlockouts
\usepackage{flushend}
\usepackage{balance}
\usepackage{cite}
\usepackage{amsmath,amssymb,amsfonts}
\usepackage{algorithmic}
\usepackage{graphicx}
\usepackage{textcomp}
\usepackage{xcolor}
\usepackage{algorithm}
\usepackage{comment}
\usepackage{tabu}
\usepackage{subfig}
\usepackage{graphicx}
\usepackage[linesnumbered,lined,boxed,commentsnumbered,algo2e]{algorithm2e}
\def\BibTeX{{\rm B\kern-.05em{\sc i\kern-.025em b}\kern-.08em
    T\kern-.1667em\lower.7ex\hbox{E}\kern-.125emX}}
\DeclareMathOperator*{\argmax}{argmax} % no space, limits  

\emergencystretch=\maxdimen
\newcommand{\blue}[1]{\textcolor[rgb]{0.00,0.00,0.00}{{#1}}}
\newcommand{\green}[1]{\textcolor[rgb]{0.00,0.00,0.00}{{#1}}}
\newcommand{\red}[1]{\textcolor[rgb]{0.00,0.00,0.00}{{#1}}}

\begin{document}

%\title{Not all Transactions Impact the Same Way: A Self-Adaptive Reinforcement Learning-Driven Load Test Agent to Tune the Workload Intelligently}

%\title{Not all Transactions are the Same: A Reinforcement Learning-Driven Load Test Agent for Intelligent Tuning of Workloads}
\title{Performance Testing Using a Smart Reinforcement Learning-Driven Test Agent}

\author{\IEEEauthorblockN{Mahshid Helali Moghadam\IEEEauthorrefmark{1}\IEEEauthorrefmark{2}, Golrokh Hamidi\IEEEauthorrefmark{2}, Markus Borg\IEEEauthorrefmark{1}, Mehrdad Saadatmand\IEEEauthorrefmark{1}, Markus Bohlin\IEEEauthorrefmark{2},\\ Björn Lisper\IEEEauthorrefmark{2}, Pasqualina Potena\IEEEauthorrefmark{1}}

\IEEEauthorblockA{\IEEEauthorrefmark{1}
\textit{RISE Research Institutes of Sweden}
, Sweden \\
\{mahshid.helali.moghadam, markus.borg, mehrdad.saadatmand, pasqualina.potena\}@ri.se}

\IEEEauthorblockA{\IEEEauthorrefmark{2}
\textit{Mälardalen University}
,Västerås, Sweden \\
\{ghi19001, markus.bohlin, bjorn.lisper\}@mdh.se}}

\IEEEoverridecommandlockouts

\IEEEpubid{\begin{minipage}{\textwidth}\ \\[9pt]
\copyright2021 IEEE. Personal use of this material is permitted. Permission from IEEE must be obtained for all other uses, in any current or future media, including reprinting/republishing this material for advertising or promotional purposes, creating new collective works, for resale or redistribution to servers or lists, or reuse of any copyrighted component of this work in other works. \hfill %--> insert the copyright option applicable from above.
\end{minipage}}

\begin{comment}
\IEEEpubid{\makebox[\columnwidth]{\copyright2021 IEEE. Personal use of this material is permitted. Permission from IEEE must be obtained for all other uses, in any current or future media, including reprinting/republishing this material for advertising or promotional purposes, creating new collective works, for resale or redistribution to servers or lists, or reuse of any copyrighted component of this work in other works.\hfill} %--> insert the copyright option applicable from above.
\hspace{\columnsep}\makebox[\columnwidth]{}}
\end{comment}

\maketitle

\IEEEpubidadjcol

\begin{abstract}
\blue{Performance testing with the aim of generating an efficient and effective workload to identify performance issues is challenging. Many of the automated approaches mainly rely on analyzing system models, source code, or extracting the usage pattern of the system during the execution. However, such information and artifacts are not always available. Moreover, all the transactions within a generated workload do not impact the performance of the system the same way, a finely tuned workload could accomplish the test objective in an efficient way. Model-free reinforcement learning is widely used for finding the optimal behavior to accomplish an objective in many decision-making problems without relying on a model of the system. This paper proposes that if the optimal policy (way) for generating test workload to meet a test objective can be learned by a test agent, then efficient test automation would be possible without relying on system models or source code. 
We present a self-adaptive reinforcement learning-driven load testing agent, RELOAD, that learns the optimal policy for test workload generation and generates an effective workload efficiently to meet the test objective. Once the agent learns the optimal policy, it can reuse the learned policy in subsequent testing activities.
Our experiments show that the proposed intelligent load test agent can accomplish the test objective with lower test cost compared to common load testing procedures, and results in higher test efficiency.}
\end{abstract}


\section{Introduction}
Performance as an important quality characteristic plays a key role in the success of software products. Performance assurance is of great importance particularly in the domains where quality assurance of both functional and non-functional aspects of system's behavior is essential. For example, enterprise applications (EAs) \cite{grinshpan2012solving} with Internet-based user interfaces such as e-commerce websites are examples of systems whose success is subject to performance assurance. EAs are often the core parts of the business organizations and their performance is a prerequisite for acceptable execution of business functions \cite{brunnert2015performance}. %Internet-based EAs may face a varying number of requests by customers meanwhile, they are required to be resilient in various execution conditions.

Performance, which is also called efficiency in classifications of quality attributes \cite{ISO/IEC,glinz2007non,chung2012non}, generally describes how well the system accomplishes its functionality.
%, mainly in terms of response time and resource consumption. 
It presents time and resource bound aspects of a system's behavior, which are indicated by some common performance metrics such as throughput, response time, and resource utilization. %Performance analysis at different stages of the software development process is a way to accomplish performance assurance. 
Performance analysis is conducted to meet the primary objectives as I) evaluating (measuring) performance metrics, II) detecting the functional problems emerging under specific execution conditions such as heavy workload, and III) detecting violations of performance requirements \cite{jiang2015survey}. %The performance analysis is often performed under both typical and stress (extreme) execution conditions. The execution condition features different aspects of the execution environment like the resource availability and the workload under which the system operates.
%Performance requirements violations often occur due to the emergence of performance bottlenecks \cite{ibidunmoye2015performance, chandola2009anomaly}. A performance bottleneck is generally defined as a system or resource component which limits the performance and hinders the system from performing as required \cite{gregg2013systems}. 

%Various application-, platform- and workload-based causes can lead to the emergence of performance bottlenecks \cite{ibidunmoye2015performance}.

Performance modeling and testing are considered common approaches to accomplish the mentioned objectives at different stages of performance analysis. %Modeling techniques using modeling notations such as queueing networks, Markov processes, and Petri nets, lead to a model of the system's behavior which is often used to measure the performance metrics \cite{cortellessa2011model,harchol2013performance,kant1992introduction}.
Although performance models \cite{cortellessa2011model,harchol2013performance,kant1992introduction} provide helpful insight into the behavior of a system, there are still many details of the implementation and the execution environment that might be ignored in the modeling \cite{denaro2004early}. Moreover, building a precise detailed model of the system behavior with regard to all the factors at play is often costly and sometimes impossible.

Performance testing as another family of techniques is supposed to meet the objectives of the performance analysis by executing the software under various realistic conditions. 
\green{\textit{\textbf{Research Challenge.}} Load testing is a type of performance testing that focuses on analyzing the performance of the system when subjected to workloads. Workload is often configured as a set of concurrent (virtual) users doing different transactions on the software under test (SUT), %(e.g., login, search, order, pay)
which often mimic the behavior of the real users of the system \cite{performanceTestingGuide}. Different transactions do not have the same impact on the performance, and generating an effective test workload in an optimal way is challenging.
Common load testing approaches such as the techniques using source code \cite{zhang2012compositional} and system model analysis (e.g., performance and UML models) \cite{gu2009search, garousi2010genetic, garousi2008traffic, costa2012generating, da2011generation}, and also use case-based \cite{draheim2006realistic, lutteroth2008modeling} and behavior-driven \cite{schulz2019behavior,ferme2018declarative, walter2016asking} techniques all mostly rely on the artifacts that are not always available during the testing. Meanwhile, in the black-box testing, in order to efficiently generate an effective workload identifying the performance effects of the transactions involved in the workload is important and still challenging.} Therefore, this paper is organized based on addressing the following question:\\ \\ \\
%on source code, system models, and behavior specifications. Nonetheless, those artifacts might not be available all the time during the testing.
%\blue{\textit{\textbf{Research Goal.}} The paper is organized based on addressing the following question:}

\textit{\textbf{Research Goal.}} \textit{How can we efficiently and adaptively generate test workload resulting in reaching the performance test objective for a SUT without relying on the source code and performance/system models?}

\textit{\textbf{Contribution.}} In this paper, we present a self-adaptive model-free reinforcement learning load testing agent (RELOAD), which learns how to generate an effective test workload efficiently without relying on the system model or source code, and is able to reuse the learned policy in further testing scenarios. \green{The test objective is defined as 
%finding performance breaking points of the SUT. A performance breaking point refers to a 
reaching a status under which a certain performance requirement gets violated.} 
%In our case reaching a performance breaking point is regarded as exceeding a certain error rate or response time threshold in received responses from the SUT.} 
%This work is a complementary system to our previously developed machine learning-assisted performance testing framework \cite{moghadam2019machine} called SaFReL. SaFReL \cite{moghadam2019machine} is a smart stress testing framework generating \textit{platform-based} stress test conditions resulting in the target performance breaking points for different software programs without access to source code or system model; in the current paper, we address the generation of \textit{workload-based} test conditions resulting in target performance breaking points.

\textit{Solution proposal.} \green{The proposed reinforcement learning-driven load testing agent
identifies the effects of different transactions involved in the workload 
%to push the system towards the performance breaking points 
and learns how to adjust the transactions 
%(e.g., number of users running each transaction) 
to meet the test objective.} It assumes two learning phases: initial and transfer learning phases. It learns the optimal policy (way) to generate an effective workload in the initial learning. Then, in the transfer learning it is able to reuse adaptively the learned policy in further testing scenarios, i.e., with different test objectives. It uses Q-learning, a model-free reinforcement learning (RL) algorithm, as the core learning with an adaptive action selection strategy to be able to reuse the learned policy in the transfer learning. RELOAD uses a well-known load test actuator, i.e., Apache JMeter \cite{ApacheJmeter}, to execute the designed workload on the SUT.

\textit{Experimental evaluation.} We present a two-fold experimental evaluation, i.e., efficiency and sensitivity analysis, of the proposed approach on a functional e-commerce web application as SUT. % which was built using widely-used open-source WooCommerce platform. %running on a local server.
\blue{In the experimental evaluation we address two main research questions which are as follows:\\
\textbf{RQ1:} How efficiently can RELOAD generate an effective test workload to meet the test objective?\\
\textbf{RQ2:} How is the efficiency of RELOAD affected by changing the learning parameters?
}

%We demonstrate how the proposed smart test agent learns to generate an effective workload efficiently and how it uses the learned policy for efficient generation of the workload in further testing scenarios.
We consider test cost saving (reduction) and compare the efficiency of RELOAD based on four configurations of the proposed learning with a random (exploratory) and a standard baseline load testing approaches. 
%in terms of test cost saving size of the generated workload (number of users) with a baseline (typical) load testing technique.
%The baseline approach treats all the transactions the same and increases the number of virtual users for all types of transactions equally. Meanwhile, the random approach, as a common black-box load testing method, chooses a transaction randomly and increases the number of generated virtual users for the selected transaction.
According to the results of the efficiency analysis, after the initial learning RELOAD generates a more accurate and finely-tuned workload to meet the test objective with around $32\%$ and $17\%$ test cost saving compared to baseline and random approaches respectively. Moreover, once it learns how to tune the transactions to reach the objective, it reuses the learned policy and keep the efficiency, i.e., preserve around $25\%$ and $13\%$ test cost saving compared to baseline and random approaches respectively, in further testing scenarios without a need to redo the learning. Lastly, we also study the behavioral sensitivity of RELOAD to the learning parameters influencing the learning mechanism.

The rest of this paper is organized as follows: Section \ref{C:Sec:: Motivation and background} discusses the motivation for applying model-free reinforcement learning to the problem and the primary concepts of RL. Section \ref{C:Sec::RL-assisted Load Testing} presents the architecture and technical details of the proposed RL-assisted load testing agent. Section \ref{sec:method} presents the research method and experiments' setup.
Section \ref{C:Sec::Result&Discussion} discusses the experimental results, answers to RQs and the threats to validity. Section \ref{C:Sec::RealtedWork} gives an overview of the related work. The conclusion and future research directions are presented in Section \ref{C:Sec::Conclusion}.     

\section{Motivation and Background} \label{C:Sec:: Motivation and background}
%\blue{A typical load testing procedure to stress a SUT towards performance breaking points involves running a load generator tool (e.g., Apache JMeter \cite{ApacheJmeter}) for a certain period of time and increasing the generated workload gradually.}
Any anomalies in the performance behavior of the system (e.g., performance requirement violation) could be mainly a consequence of emerging bottlenecks at the level of platform or application \cite{ibidunmoye2015performance, chandola2009anomaly}. A bottleneck can make the system fail or not perform as required, and can happen due to the full utilization of the component capacity, exceeding a usage threshold or occurrence of contention \cite{gregg2013systems}.

Possible defects in source code or architecture and some issues related to platform resources could be often the root causes of the emergence of bottlenecks. %under different workload conditions.
Moreover, all transactions do not have the same effect on the performance and some of them are more critical to lead to the emergence of performance bottlenecks. 
%The application-based causes might vary, e.g., during the continuous integration/continuous delivery (CI/CD) process, and the platform-based ones might also vary during the execution and act differently under different workload conditions.
Therefore, due to the existing interplay between the involved factors, drawing a detailed model expressing the performance behavior of the system, is not easily possible. This issue makes room for model-free machine learning techniques, such as model-free reinforcement learning (RL) \cite{sutton2018reinforcement} to play an interesting role in addressing the related challenges, in particular from testing perspective. RL algorithms are mainly intended to address decision-making problems and have been widely used to build self-adaptive intelligent systems. 

\blue{In model-free RL the intelligent agent can learn an optimal behavior to achieve an intended objective based on the interaction with the environment (i.e., the system under test in this problem) without access to the source code or a model of SUT.}
%It is able to act to achieve the intended objective in an adaptive way to varying conditions.
\blue{Furthermore, the agent is able to store the gained knowledge and reuse the learned behavior in further potential testing situations such as regression testing or testing of SUT with regard to different test objectives. Model-free RL algorithms are not intended to build or learn a model of the environment. Instead, they learn optimal behavior to accomplish the objective through various episodes of interaction with the environment. They are apt for the problems where the model (i.e., the dynamics) of the environment is unknown or costly to be built, but the experience of interaction with the environment can be sampled and used.} 
%Based on these capabilities of model-free RL, in this paper, we aim at generating the workload-based test conditions resulting in reaching a target response time and error rate and response time threshold in responses of SUT  without access to source code or system models.  

\subsection{Reinforcement Learning} \label{Sec: RL}
%Reinforcement learning is a learning technique that is mainly based on the interaction between an agent and the environment (system) of the problem. It has been widely used to build self-adaptive intelligent systems.
%The principles of RL is as follows:

Using RL, the agent learns the optimal behavior to meet the objective through being rewarded or punished in the interaction with the environment. At each step of the interaction, the agent observes the state of the system. It takes one possible action. The system undergoes changes upon actions. Then, the agent receives a reward signal showing how good the action was to direct the agent towards accomplishing the objective. The overall goal of the agent is formulated in terms of maximizing the cumulative long-term reward. The agent decides how to behave at each step of the interaction and based on optimizing the long-term received reward, learns the optimal behavior function which is called \textit{optimal policy}. % In a model-free RL, the agent generally learns an optimal policy to achieve the goal.
The agent uses an action selection strategy to interact with and apply actions to the system. The action selection is often based on trying the available actions, i.e., exploration of the action space, or relying on the learned policy which leads to selecting highly valued actions, i.e., exploitation of the gained knowledge.
\section{RELOAD Test Agent for Optimal Test Workload Generation}\label{C:Sec::RL-assisted Load Testing}
In this section, we present an overview of the architecture of our proposed RL-driven load testing agent, RELOAD, and describe the technical details of the learning procedure. %In summary, the proposed self-adaptive intelligent load testing agent features as follows:\\
%\textit{How it addresses the problem.} The proposed load test agent learns how to tune the transactions in the test workload and generate an effective workload efficiently to accomplish the test objective, without access to system model or source code.\\
%It learns the optimal policy to generate the proper load resulting in performance issues of the SUT, before the saturation point, under different execution environments .\\

\textit{How it learns.} 
It assumes two phases of learning, i.e., initial and transfer learning. During the initial learning, the test agent learns the optimal policy to generate an effective workload to accomplish the test objective. %i.e., reaching the intended performance breaking point. 
%to find the performance issues which can emerge before system saturation. 
During the transfer learning, the learned policy is reused in further potential testing scenarios, e.g., regression testing scenarios or testing with regard to different test objectives. In the transfer learning phase, the agent also still continues with the learning to keep the policy updated.
%execution of SUT on different platforms,
 
\begin{figure}[ht]
  \centering
  \includegraphics[width=.98\columnwidth, height=5cm]{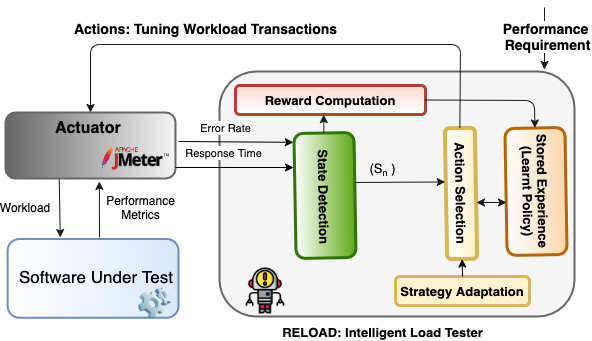}
  \caption{RELOAD, an RL-driven load testing agent}
  \label{fig: architecture of load runner}
\end{figure}

We use Q-learning \cite{sutton2018reinforcement}, a model-free RL algorithm, as the core learning technique. Fig. \ref{fig: architecture of load runner} shows the architecture of RELOAD. The main constituent parts of each learning step in RELOAD are \textit{detecting state}, \textit{taking actions} and \textit{computing reward} (See Section \ref{Sec: RL}). We have formulated these parts in RELOAD as follows:\\

\begin{figure*}[t]
  \centering
  \includegraphics[width=.8\textwidth, height=4cm]{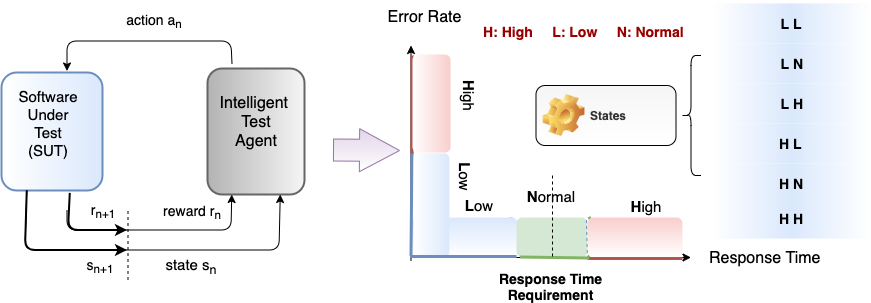}
  \caption{States of SUT in the proposed RL-based load testing }
  \label{fig: States}
\end{figure*}

\textbf{State Detection.} Average response time and error rate, as two performance metrics, are used to indicate the \green{performance} state of the SUT. The values of these performance metrics are classified under a number of discrete classes, which are described as \textit{Low}, \textit{Normal} and \textit{High} for response time and \textit{Low} and \textit{High} for error rate. \green{The threshold (boundary) values for defining these classes are selected empirically and could be updated based on the requirements.} The combinations of these classes form the discrete classes for the state of the system, as shown in Fig. \ref{fig: States}. \green{Actually, different transactions do not have the same impact on the performance of the SUT, and test workloads with different configurations, i.e., in terms of constituent transactions, might lead the SUT to different performance states.} The agent fetches these metrics from the test actuator at each learning step and identifies the state of the SUT.
%Since response time and error rate are correlated substantially. Therefore, the combinations of (High, Low) and (High, Normal) for error rate and response time respectively do not occur and are not valid.

\textbf{Actions.} At each learning step, the test agent takes one action after detecting the state of the SUT. We define the actions as adjusting the load of constituent transactions in the workload, in terms of numbers of virtual users running each transaction. %The list of the involved transactions is definitely application-specific.
Table \ref{table: transactions} presents the list of transactions for the SUT in our case study, which is an e-commerce web application.\\
%Before starting the learning, the set of involved requests in each transaction is set through recording a simple usage of the SUT by a user. Many of the load generator tools such as Apache JMeter, which actually acts as a test actuator in our case study, have the functionality of recording the requests involved in a transaction which is run on the SUT. %to make an executable test plan for testing the SUT.

\red{Each transaction involves a certain function together with its functional dependencies. For example, transaction \textit{Add to cart} involves performing \textit{login, accessing the search page, and selecting the product} as well, since all those functions are prerequisites for function \textit{Add to cart}. 
%Each transaction might also have some functional dependency, i.e., some certain operations are required to be done before a particular transaction. For example, the functional prerequisites for transaction \textit{Add to cart}, are \textit{login, accessing the search page, and selecting the product}.
Therefore, the function in each transaction of the workload is considered together with its functional dependencies.} Then, the set of actions for the test agent is defined as follows:
\begin{comment}
\begin{equation} \label{eq:action}
ActionList = \{\cup\ action_k,\ 1 \leq k \leq |\textit{List of Transactions}| \}
\end{equation}
\begin{equation} \label{eq:action2}
\begin{split}
action_k=\{\cup\ (W_n^{T_j})\ |\ & W_n^{T_j} = W_{n-1}^{T_j},\ \ if\  j\ne\ k,\\
           & W_n^{T_j}= W_{n-1}^{T_j}\ +\ \frac{W_{n-1}^{T_j}}{3},\ if\ j=k,\\
           & T_j \in \textit{List of transactions},\\
           & 1 \leq j \leq |\textit{List of Transactions}|\}
\end{split}
\end{equation}
\end{comment}
\begin{equation} \label{eq:action}
ActionList = \{\cup\ action_k,\ 1 \leq k \leq |\textit{List of Transactions}| \}
\end{equation}
\begin{equation} \label{eq:action2}
\begin{split}
action_k: \{& W_n^{T_j} = W_{n-1}^{T_j},\ \ for\  j\ne\ k,\\
           & W_n^{T_j}= W_{n-1}^{T_j}\ +\ \frac{W_{n-1}^{T_j}}{3},\ for\ j=k,\\
           & T_j \in \textit{List of transactions},\\
           & 1 \leq j \leq |\textit{List of Transactions}|\}
\end{split}
\end{equation}
where $T_j$ indicates a transaction of the SUT. $W_n^{T_j}$ indicates the load of transaction $T_j$ at time step $n$, i.e., the number of users running this transaction. % In each action, running the transaction implies running its prerequisite transactions as well.
After the agent decides on an action, a test plan is generated by the agent, and then is executed on the SUT by the test actuator, i.e., Apache JMeter. %The batch of requests from the client to the server.

%the operations which are prerequisite to the transaction. For example, running transaction $T_{add to cart}$, includes the operations \textit{Login page}, \textit{Login}, \textit{Search page}, \textit{Select product} and, \textit{Add to cart}.
%The modified workload in each action runs for a certain period which can be defined empirically.
%and in proportional to the total approximate configured time for the testing.
%action_k=\{(T_k^{Wl}\ +\ \frac{T_k^{Wl}}{4})\ |\ T_k \in \textit{List of transactions}\}
\begin{table}[h!]
\caption{List of transactions for the SUT}
\begin{center}
\begin{tabular}{|p{2.5 cm}|p{4.5cm}|}
 \hline
 \textbf{\textit{Operation}} & \textbf{\textit{Description}} \\
 \hline
 Home & Access to home page \\
 \hline
 Sign up page & Access to Sign up page\\
 \hline
 Sign up & Register and add a new user\\
 \hline
 Login page & Access to login page \\
 \hline
 Login & Sign in at the system\\
 \hline
 Search page & Access to search page\\
 \hline
 Select product & See the details of the selected product\\
 \hline
 Add to cart & Add the selected product to the cart\\
 \hline
 Payment & Access to payment page\\
 \hline
 Confirm & Confirm the order (payment) \\
 \hline
 Log out & Log out \\
 \hline
\end{tabular}
\label{table: transactions}
\end{center}
\end{table}

\textbf{Reward Signal.} After taking the selected action and running the tuned workload, the test agent receives a reward signal which shows how effective the applied action was in leading the test agent to reaching the test objective. %i.e., the intended performance breaking point. 
%finding more performance issues of SUT before reaching the saturation point of the environment.
We define a function to represent the reward signal as follows:

\begin{equation} \label{eq:reward}
 R_n = (\frac{RT_n}{RT_{threshold}})^2+(\frac{ER_n}{ER_{threshold}})^2
\end{equation}

where $R_n$, $RT_n$, and $ER_n$ indicate the reward, the average response time, and the average error rate respectively, in step $n$. Also $RT_{threshold}$ and $ER_{threshold}$ are the response time and error rate thresholds related to the test objective.
%performance breaking point.

\textbf{Learning Procedure.} In RL, the agent is intended to learn the optimal policy to accomplish the objective of the problem. The policy determines which action to be taken by the agent, given a certain state. The key idea for finding the optimal policy is the use of an iterative policy iteration process at each step of the learning, which consists of policy evaluation and policy improvement. At each step of the interaction, the agent performs both evaluation and improvement. First, it evaluates the policy which it follows, %(e.g., based on estimating the value function),
then it tries to improve it through a greedy approach (e.g., $\varepsilon$-greedy). Finally, this process will converge on the optimal policy.    
In model-free RL, there are generally two approaches to realize this: learning the policy directly and indirectly. In the Q-learning algorithm, the agent learns an optimal value function, i.e., an action-value function \(Q^*(s,a)\), from which the optimal policy can be obtained. The optimal action-value function, \(Q^*(s,a)\), gives the expected long-term return, given state $s$, taking an arbitrary action $a$, and then following the optimal policy. It is presented as follows:
\begin{equation} \label{eq:optimal action-value}
Q^*(s,a)= \argmax\limits_\pi E^\pi [q_n | s_n=s,a_n=a]
\end{equation}
\begin{equation} \label{eq:discounted return}
q_n=\sum_{k=0}^{\infty} \gamma^k R_{n+k+1}
\end{equation}
where \(\gamma\) is a discount factor for future rewards and $q_n$ is the long-term return in terms of cumulative discounted reward.
In general, the optimal policy selects the action maximizing the expected return given starting from state $s$. Moreover, according to the definition of \(Q^*(s,a)\), given $Q^*$, the optimal action for state $s$, $a^*(s)$, is obtained as:
\begin{equation} \label{eq: optimal action}
a^*(s)= \argmax\limits_{a^{'}} Q^*(s,a^{'})
\end{equation}	
In order to obtain the optimal policy, Q-values are stored (e.g., in a Q-table or a neural network) and considered the experience of the agent. During the learning, the Q-values are updated incrementally according to Eq. \ref{eq:Q updating}:
\begin{equation} \label{eq:Q updating}
Q(s_n,a_n)=(1-\alpha) Q(s_n,a_n)+ \alpha[R_{n+1}+ \gamma \max\limits_{a^{'}} Q(s_{n+1},a^{'})]
\end{equation}
where \(\alpha\), \(0 \leq \alpha \leq 1\) adjusts the rate of learning which controls the impact of new Q-values on the previous ones.

%\textit{Learning phase.} 
%The proposed Rl-based load runner assumes two phases of learning, i.e. \textit{initial} and \textit{transfer} learning. 
In this study the research problem, i.e., generating an effective workload to meet an intended test objective, is regarded as a sequential decision-making problem. Model-free RL is proposed as a beneficial learning solution to this problem since the SUT (environment) and execution platform are supposed to be initially unknown to the test agent. Then, in the proposed model-free RL-driven solution, 
%the problem is treated as a Markov Decision Process (MDP), and 
the agent finds (learns) the optimal policy to generate an effective workload to accomplish the test objective through a built-in iterative policy evaluation-improvement process. 
%One of the basic paradigms in solving an MPD is finding an optimal policy through policy and value iterations techniques. The value iteration technique has been used in this study. 
Algorithms \ref{Algorithm: Rl-driven load testing} and \ref{Algorithm: Q-Learning} present the procedure of the learning in the proposed RL-driven load testing agent.

In model-free RL, $\varepsilon$-greedy is a well-known method for action selection, when RL is used to find the optimal policy in a decision-making problem. It guarantees the sufficient continual exploration required for finding the optimal policy, and meanwhile provides a proper trade-off between exploration of the state-action space and exploitation of the learned value function.   
In $\varepsilon$-greedy, the value of $\varepsilon$ adjusts the degree of exploration versus exploitation, as it leads the agent to select a high-value action based on the learned value function with probability (1-$\varepsilon$) or a random possible action with probability $\varepsilon$, given a certain state. In addition to Q-learning, we also implemented RELOAD with DQN \cite{mnih2015human}, which is a combination of Q-learning and deep neural networks and suits the large scale problems where due to the big number of states and actions using tabular methods (i.e., Q-table) is not practical.

\begin{algorithm}[h]
\SetAlgoLined
\begin{flushleft}
\caption{Adaptive Reinforcement Learning-Driven load Testing}\label{Algorithm: Rl-driven load testing}
%\begin{algorithm2e}
%\begin{justifying}
\textbf{Required:} \(\mathbb{S}, \mathbb{A}, \alpha, \gamma\);\\
\textrm{Initialize Q-values,\  \(Q(s,a)= 0\ \forall s \in \mathbb{S},\  \forall a \in \mathbb{A}\ \textrm{and}\ \varepsilon=\upsilon\ ,0 < \upsilon <1\)};\\
%\textrm{Initial Learning};\\ 
%Initial Learning:\\
\While{ Not (initial convergence reached)}{
        Learning\_Episode (with initial action selection strategy, e.g., $\varepsilon$-greedy, initialized $\varepsilon$)\;
    }

\textrm{Store the learned policy};\\
%\textrm{Start the transfer learning phase};\\
%\textbf{Transfer Learning:}\\
\textrm{Adapt the action selection strategy to transfer learning, i.e., tune parameter $\varepsilon$ in $\varepsilon$-greedy};\\
\While{true}{
Learning\_Episode with adapted strategy (e.g., new value of $\varepsilon$);\\
}
%\end{justifying}
%\end{algorithm2e}
\end{flushleft}
\end{algorithm}

\begin{algorithm}[h]
\SetAlgoLined
\begin{flushleft}
\caption{Learning\_Episode}\label{Algorithm: Q-Learning}
%\begin{algorithm2e}
%\begin{justifying}
\Repeat{meeting the stopping criteria (reaching the test objective) }{
        \textrm{1. Detect the state \((S_n)\) of the SUT};\\ 
        \textrm{2. Select an action (See Eq. \ref{eq:action}) according to the action selection strategy, e.g., $\varepsilon$-greedy: select $a_n= \argmax_{a\in \mathbb{A}} Q(s_n,a)$ with probability (1-$\varepsilon$) or a random $a_k$, $a_k \in \mathbb{A}$ with probability $\varepsilon$};\\
        3. Take the selected action: Tune the workload and run the modified workload on the SUT;\\
        4. Detect the new state $(S_{n+1})$ of the SUT;\\
        5. Compute the reward, $R_{n+1}$;\\
        6. Update the Q-value of the pair of previous state and taken action\\ 
        $Q(s_n,a_n)=(1-\alpha) Q(s_n,a_n)+ \alpha[R_{n+1}+ \gamma \max\limits_{a^{'}} Q(s_{n+1},a^{'})]$\;	
    }
%\end{justifying}
%\end{algorithm2e}
\end{flushleft}
\end{algorithm}

\section{Method} \label{sec:method}
We perform empirical evaluations of RELOAD by running experiments on a mature open-source software, an e-commerce web application. Our target SUT is based on the widely-used WooCommerce platform and deployed using XAMPP on an Apache web server with PHP 7.4.13 and MariaDB 10.4.17. The experiments' environment consists of two virtual machines (VMs), as one of them hosts the SUT and the other one runs the load testing agent together with the test actuator. Each VM has 2 CPUs at $3.1$GHz, 8GB of RAM, and Linux Ubuntu 16.04. %  hosting server with one shared CPU and up to 2GB shared RAM. The SUT supports 11 common use cases that have been described in Table \ref{table: transactions}.
We use Apache JMeter 5.2.1 as an actuator to execute the test workload on the SUT. 

We design a series of experiments to assess the efficiency and sensitivity of RELOAD. The experiments investigate how different learning configurations (setups) affect the outcome of RELOAD. For comparative purposes, we also report results from random (exploratory) testing and a standard (naive) testing baseline. For all experimental runs, we translate differences in the number of generated concurrent virtual users to reduced testing costs.

%purposes of the experiments are to assess how efficiently RELOAD, i.e., with regard to various configurations of the learning, can perform in reaching test objectives and how sensitive its efficiency is to the learning parameters. We compare its efficiency with a random (exploratory) and a baseline load testing approaches in different testing scenarios and show how it benefits the learning and leads to reduced testing costs.  

\begin{figure}
  \centering
  \includegraphics[height=2.5cm, width=.90\columnwidth]{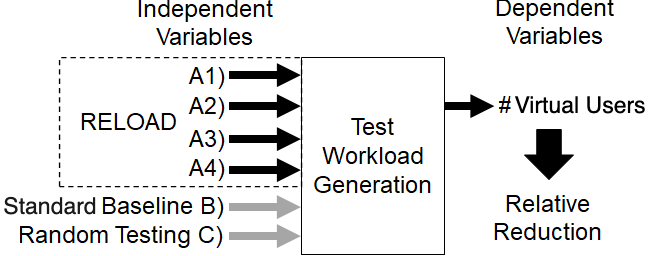}
  \caption{Overview of the experimental setup.}
  \label{fig:expSetup}
\end{figure}

Figure~\ref{fig:expSetup} shows an overview of the experimental setup. The Dependent Variable (DV) in all experimental runs is the 
number of generated virtual users.
%episodes until convergence.
The Independent Variable (IV) defining different experimental runs is the test generation technique. We explore six discrete levels of the IV: A1) RELOAD with $\varepsilon = 0.2$, A2) RELOAD with $\varepsilon = 0.5$, A3) RELOAD with decaying $\varepsilon$, A4) RELOAD with DQN; B) Standard Baseline; C) Random Testing. In A1)-A3) RELOAD is based on Q-learning together with $\varepsilon$-greedy with different values for $\varepsilon$.

Section~\ref{C:Sec::RL-assisted Load Testing} describes the details of the RELOAD configurations in A1)-A4). The Standard Baseline (B) applies an initial workload that contains all the transactions with the \textit{same} number of users per each transaction, then increases the number of users in fixed steps by $33\%$ until accomplishing the test objective.
In Random Testing (C), a transaction is chosen randomly at each step, and then the number of virtual users allocated for the selected transaction is increased by $33\%$. The process is repeated until the test objective has been met.

The experimental runs corresponding to the six test generation techniques (i.e., values of the IV) are executed the same number of times, i.e., the same number of \textit{episodes}. In RL each learning \textit{episode} constitutes one complete sequence of states and actions in RL till reaching the objective (i.e., equivalent to one epoch in supervised learning). The agents' properties including the value function and policy are updated gradually over the learning episodes. Despite the lack of learning in the Baseline and Random testing, we refer to one complete execution for those techniques as an episode too. %The test objective is to find a performance breaking point under which 1) the response time of the SUT exceeds $1,500 ms$ or 2) the error rate in the received responses exceeds $20\%$.   

%while our intelligent tester agent designs the test workload.
%(i.e., the intelligent test agent, RELOAD, creates the test plan workload and Apache JMeter executes it. 

% In our evaluation setup, we set the response time requirement and the boundary value between the error rate states (See Fig. \ref{fig: States}) to $1300 ms$ and $0.05$ respectively, which are determined empirically based on the SUT requirements.

%We evaluate the proposed approach through two analysis scenarios, i.e efficiency, and sensitivity analysis. 

%In the former, we study the efficiency of the proposed approach in generating a target test workload to meet the intended objective.

%as the number of episodes in the learning-based approach, and meanwhile the efficiency of the learning over the learning iterations (episodes) is shown against benchmark approaches. A learning episode is a one-time execution of the learning that leads to reaching the test objective. The agents' properties including the value function and policy are updated over the learning episodes.

In the \textbf{efficiency analysis}, we report results corresponding to the two learning phases of RELOAD. First, we analyze the initial learning. Second, we study how efficiently RELOAD performs during the transfer learning, i.e., when the agent reuses learned policies in new similar testing scenarios.  

In the \textbf{sensitivity analysis}, we investigate the performance sensitivity of RELOAD to two learning hyperparameters, i.e., the \textit{learning rate} $\alpha$ and the \textit{discount factor} $\gamma$. We explore the two hyperparameters by changing one parameter while keeping the other one constant. As the sensitivity analysis followed the efficiency study, we based the design on our empirical observations at that point.

In the efficiency experiments, we use baseline values of $\alpha=0.5$ and $\gamma=0.5$. In the sensitivity experiments, we conduct four experimental runs to analyze the sensitivity of RELOAD. First, we set $\alpha$ to $0.1$ and $decaying$ values while keeping the value of $\gamma$ fixed at $0.5$. Second, we set the $\gamma$ to $0.1$ and $0.9$, while fixing $\alpha$ at $0.5$.

\section{Results and Discussion}\label{C:Sec::Result&Discussion}
This section presents our experimental results, answers the RQs, and discusses the main threats to validity.  

\subsection{Experimental Results}
\textbf{\textit{Efficiency Analysis.}} 
%the stopping condition of the load testing in both RL-assisted and the typical process was set as reaching a certain error rate threshold which was $50\%$ in our early-stage experimentation.
%In the efficiency analysis experiments, we set the learning rate and discount factor to $0.5$ and $0.5$ respectively. Moreover, the performance-related status of the hosting server, i.e., the resource availability, remains unknown to the intelligent load tester.
%Trials indicate the number of steps that the workload was modified (i.e., the taken actions) by the intelligent agent.
%Each experiment starts with running an initial workload, and trials indicate the number of steps that the workload was modified (taken actions) by the intelligent agent. We run each experiment 10 times. 
%varies highly due to other co-located applications and background running processes, while the intelligent load tester does not know and observe it. 
\textit{Initial Learning}. 
%First, we analyze the efficiency of the RL-assisted load testing after initial learning in terms of the size of the generated workload meeting the intended error rate and response time threshold compared to the baseline procedure (See figure \ref{fig:efficiency}).
To see how it works during the initial learning, we compare the efficiency of RELOAD for the learning configurations A1)-A4) (i.e., $\varepsilon=0.2$, $0.5$, decaying $\varepsilon$, and DQN) with the Standard Baseline and Random Testing. In particular, we are interested in studying the behavior of RELOAD after the initial convergence in comparison with other approaches. The convergence happens after around $30$ episodes in Q-learning with $\varepsilon$-greedy (A1-A3) and in some episodes later for the DQN configuration (A4), i.e., after roughly $37$ episodes. We consider the performance of the learning-based approach during the last $10$ episodes after the convergence. We also run the Standard Baseline (B) and the Random Testing (C) $40$ episodes. 
%We run RELOAD for $40$ episodes and subsequently run the other approaches $40$ times as well.
The test objective is reaching a performance status under which 1) the response time of the SUT exceeds $1,500 ms$ or 2) the error rate in the received responses exceeds $20\%$.

\begin{figure*}[h]
\centering
\subfloat[RELOAD with Q-learning, $\varepsilon$-greedy, $\varepsilon = 0.2$]{
    \label{final_workload_epsilon_0.2}
    \framebox{\includegraphics[width=0.48\linewidth, height=4cm]{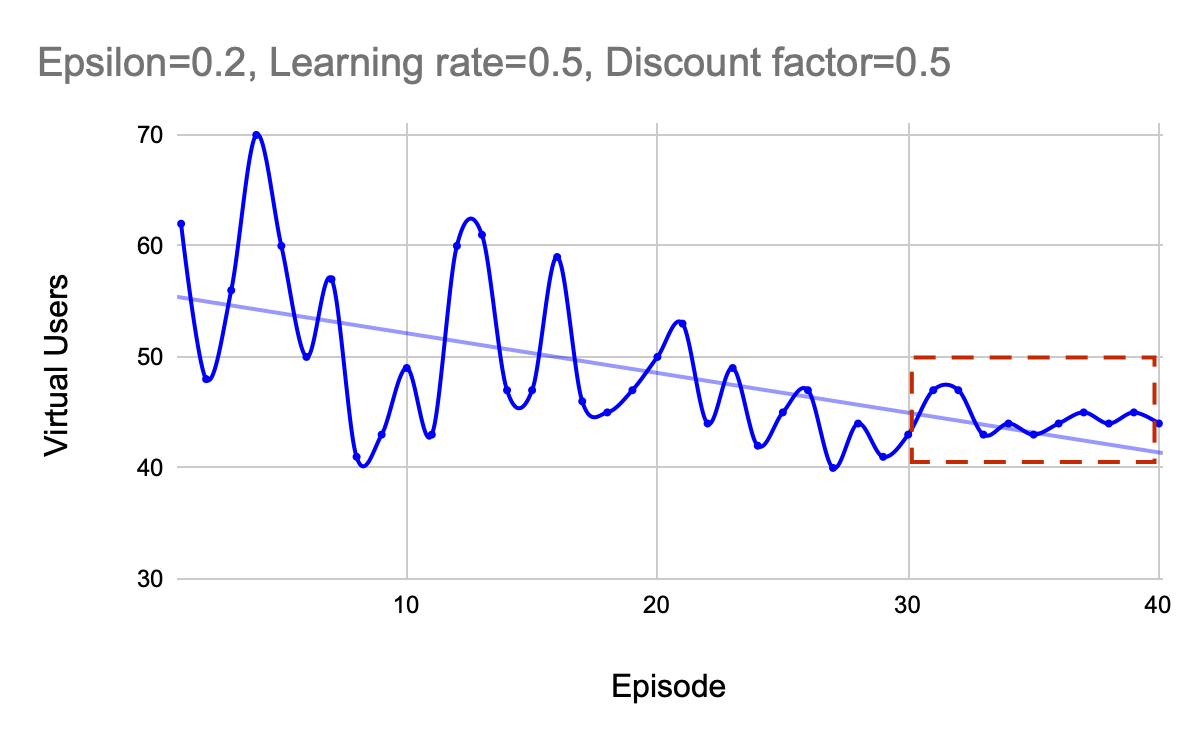}} } 
\subfloat[RELOAD with Q-learning, $\varepsilon$-greedy, $\varepsilon = 0.5$]{
    \label{final_workload_epsilon_0.5}
    \framebox{\includegraphics[width=0.48\linewidth, height=4cm]{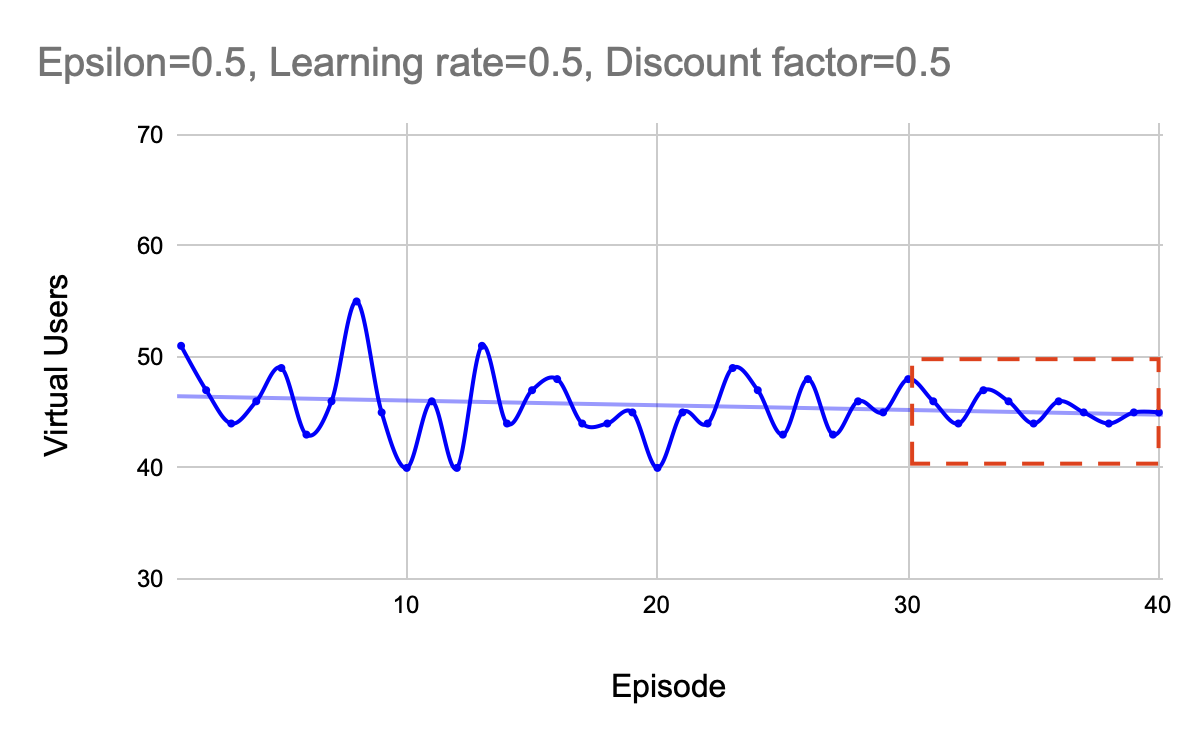}}}\\
\subfloat[RELOAD with Q-learning, $\varepsilon$-greedy, decaying $\varepsilon$]{
    \label{final_workload_decaying_epsilon}
    \framebox{\includegraphics[width=0.48\linewidth, height=4cm]{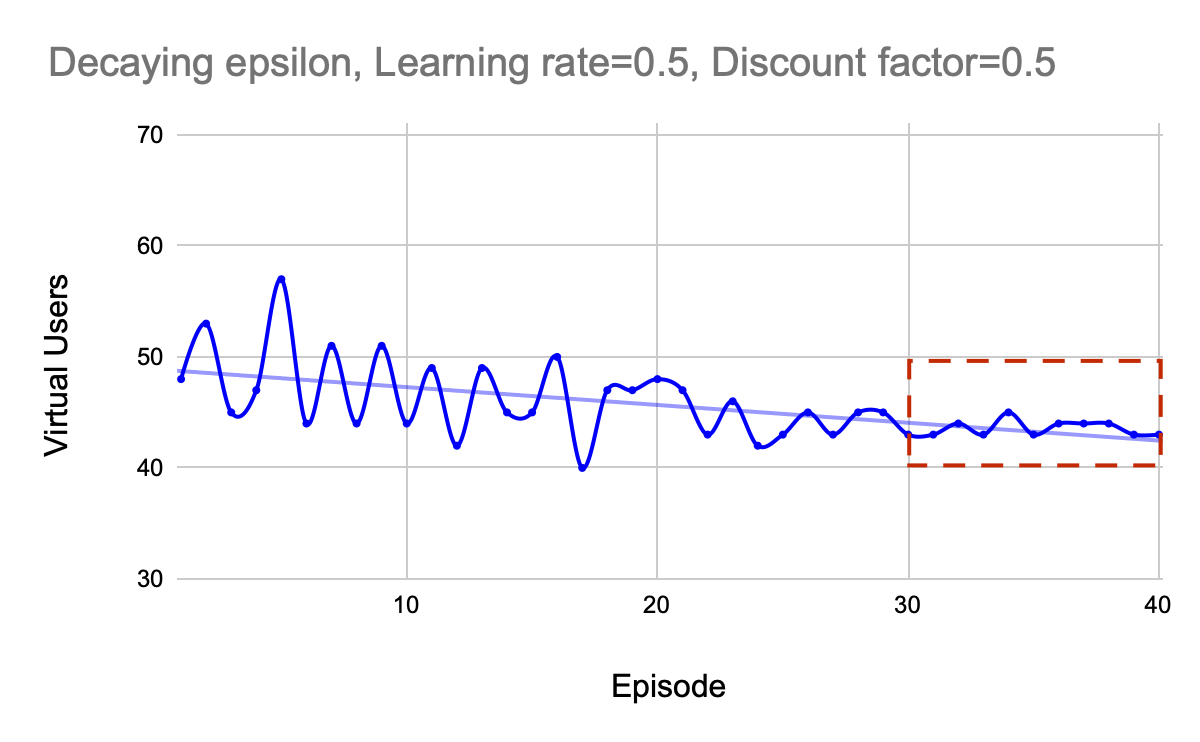}} } 
\subfloat[RELOAD with DQN setup]{
   \label{final_workload_DQN}
   \framebox{\includegraphics[width=0.48\linewidth, height=4cm]{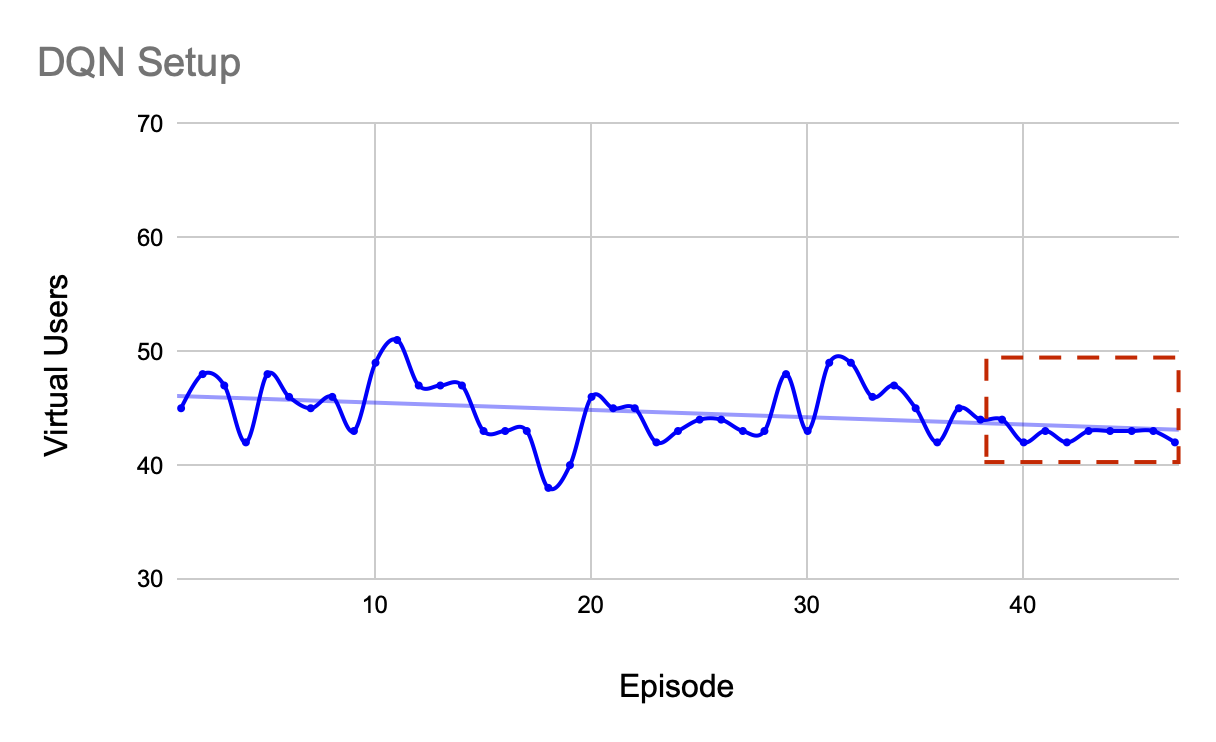}} }\newline
\subfloat[Baseline Approach]{
    \label{final_workload_baseline}
    \framebox{\includegraphics[width=0.48\linewidth, height=4cm]{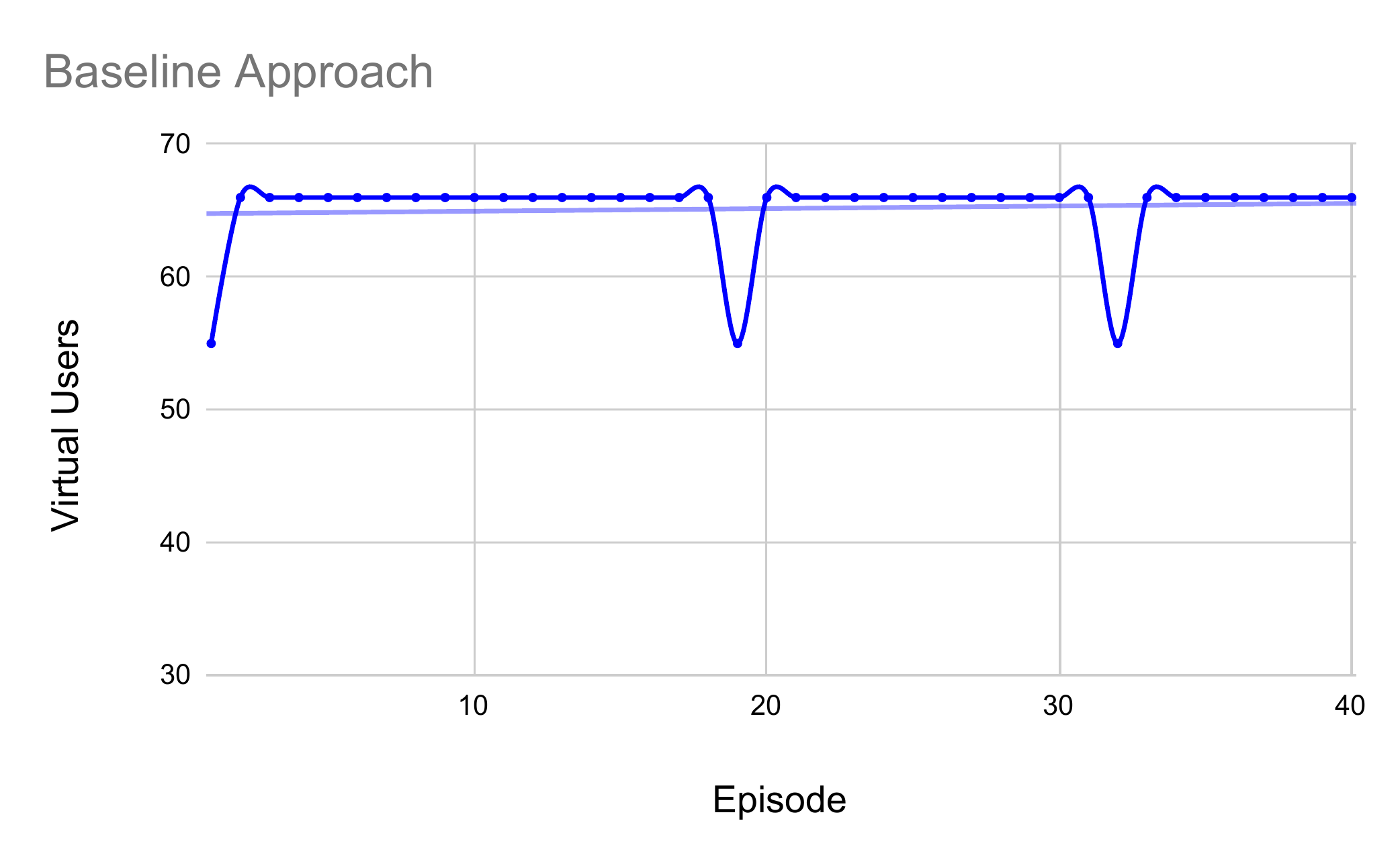}}}
\subfloat[Random Approach]{
    \label{final_workload_Random}
    \framebox{\includegraphics[width=0.48\linewidth, height=4cm]{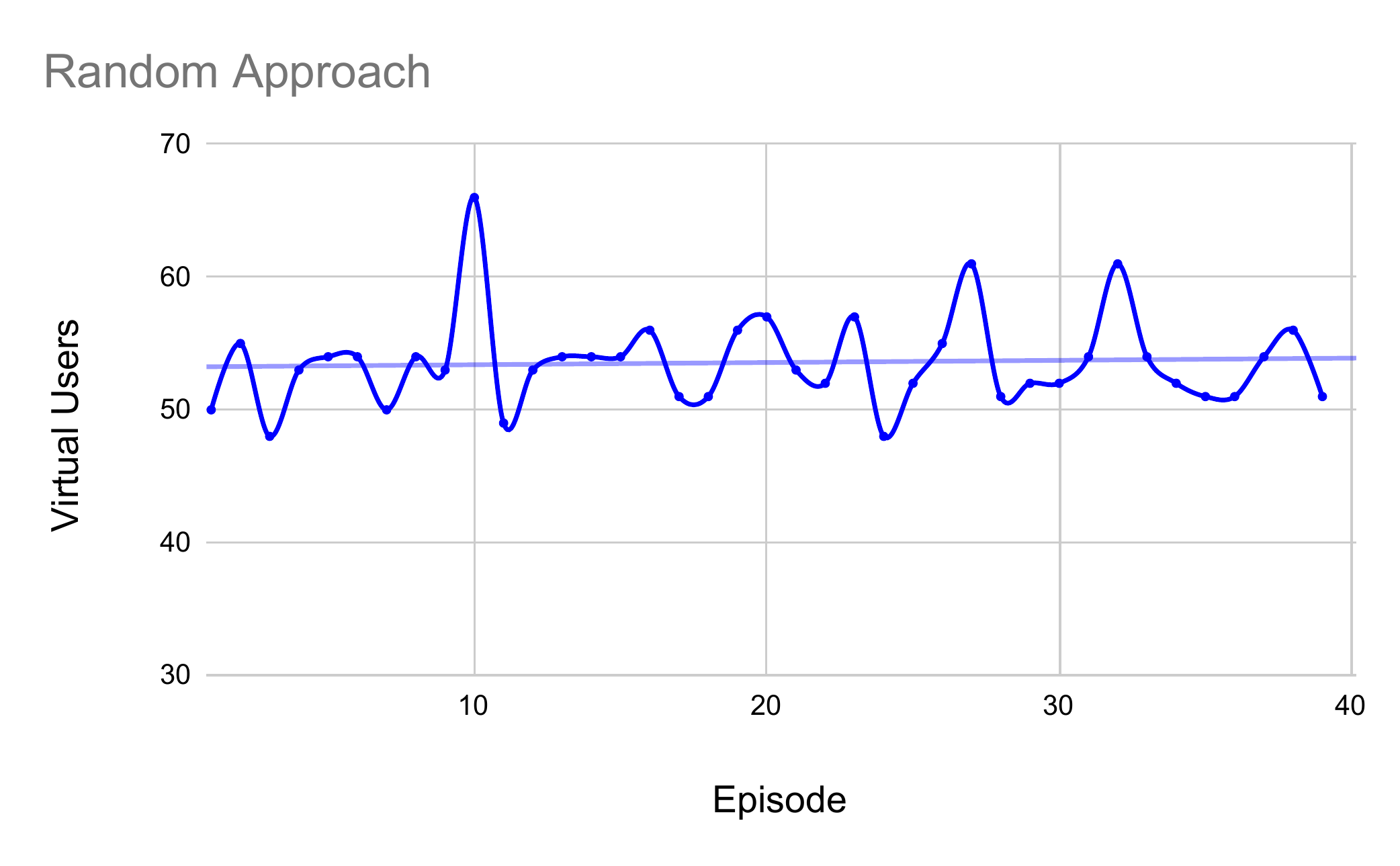}}}
\caption{Test efficiency of RELOAD (initial learning for configurations A1-A4), Standard Baseline, and Random Testing}
\label{fig:efficiency}
\end{figure*}

Fig. \ref{fig:efficiency} shows the number of generated virtual users in all approaches to produce an effective workload accomplishing the test objective.
%in the baseline approach and the RL-assisted approach regarding using $\varepsilon$-greedy with different values of $\varepsilon$. 
%As shown in Fig. \ref{fig:efficiency}, overall, the learning approach after a couple of episodes reaches the test objective with a smaller and more fine-grained workload than the baseline approach, which subsequently results in lower testing cost and time. 
Table \ref{table:Initial learning-average efficiency} presents the resulting \blue{test cost saving} at the last 10 episodes in RELOAD, i.e., the last 10 episodes show the behavior of the RL approach when it has almost achieved an initial convergence. We proceed by discussing the performance of RELOAD using the four configurations A1)-A4).

%settings of $\varepsilon$-greedy 
\textit{Q-learning with $\varepsilon$-greedy.} Using $\varepsilon = 0.2$ (A1) makes the agent mainly rely on the stored experience rather than exploring new actions. It might slow down the learning convergence in a varying environment in which more exploration is needed. This issue is observable in terms of high spikes in Fig. \ref{final_workload_epsilon_0.2}.
%Despite the spikes, the figure shows that the workload size converges over the learning episodes. Therefore, we can conclude that the agent has learned the optimal policy.\\ 
The configuration \textit{$\varepsilon = 0.5$.} (A2) provides an equal likelihood for the exploitation of the learned policy and the exploration of new actions. 
%In our experiments, it gives slightly less efficiency than $\varepsilon = 0.2$ strategy in terms of the size of the generated workload.\\
The \textit{decaying $\varepsilon$} setting (A3) decreases $\varepsilon$ gradually over the learning episodes. It makes the agent explore new actions mainly during the early episodes of the learning and do more exploitation of the learned policy in the later episodes. The efficiency of the three configurations A1)-A3) are comparable, i.e., they converge roughly on the same number of virtual users needed to meet the test objective.
\textit{DQN setup.} DQN (A4) is an extension of Q-learning that uses a deep neural network as a function approximator instead of a Q-table to approximate the Q-values. In this experiment, the A4 obtain roughly the same efficiency as Q-learning with $\varepsilon$-greedy (A1-A3). This is also in line with previous works on the use of Q-learning for performance assurance purposes \cite{ibidunmoye2017adaptive}, i.e., for problems that are not high-dimensional and satisfy the required conditions for Q-learning convergence, it is possible to obtain desired results using Q-learning with a carefully selected configuration. For the transfer learning part, we proceed with the A3 configuration.

\begin{figure*}
\centering
\subfloat[RELOAD in the Transfer Learning]{
    \label{transfer-learning-Qlearning}
    \framebox{\includegraphics[width=0.31\linewidth, height=4cm]{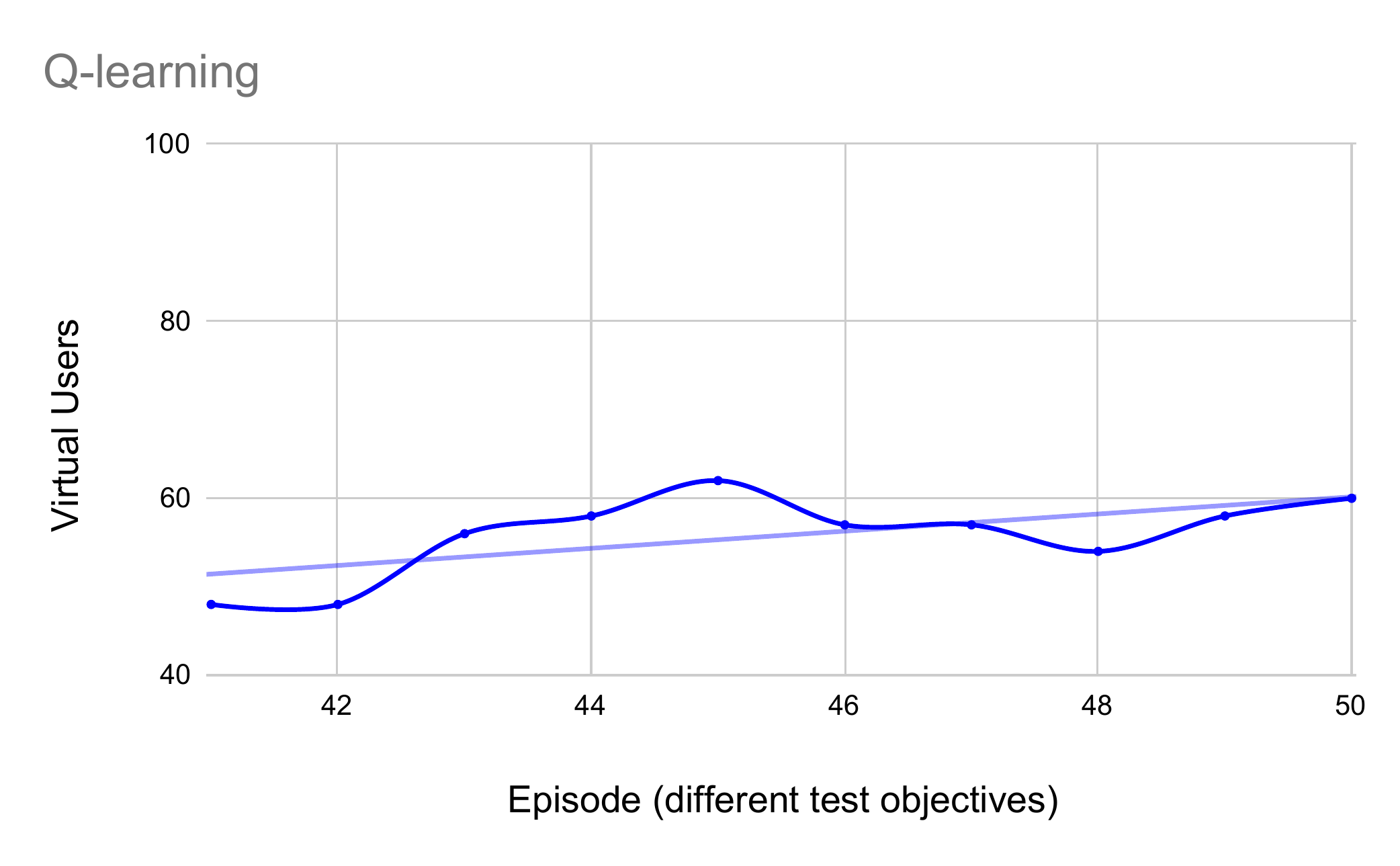}} } 
\subfloat[Standard Baseline]{
    \label{transfer-learning-Baseline}
    \framebox{\includegraphics[width=0.31\linewidth, height=4cm]{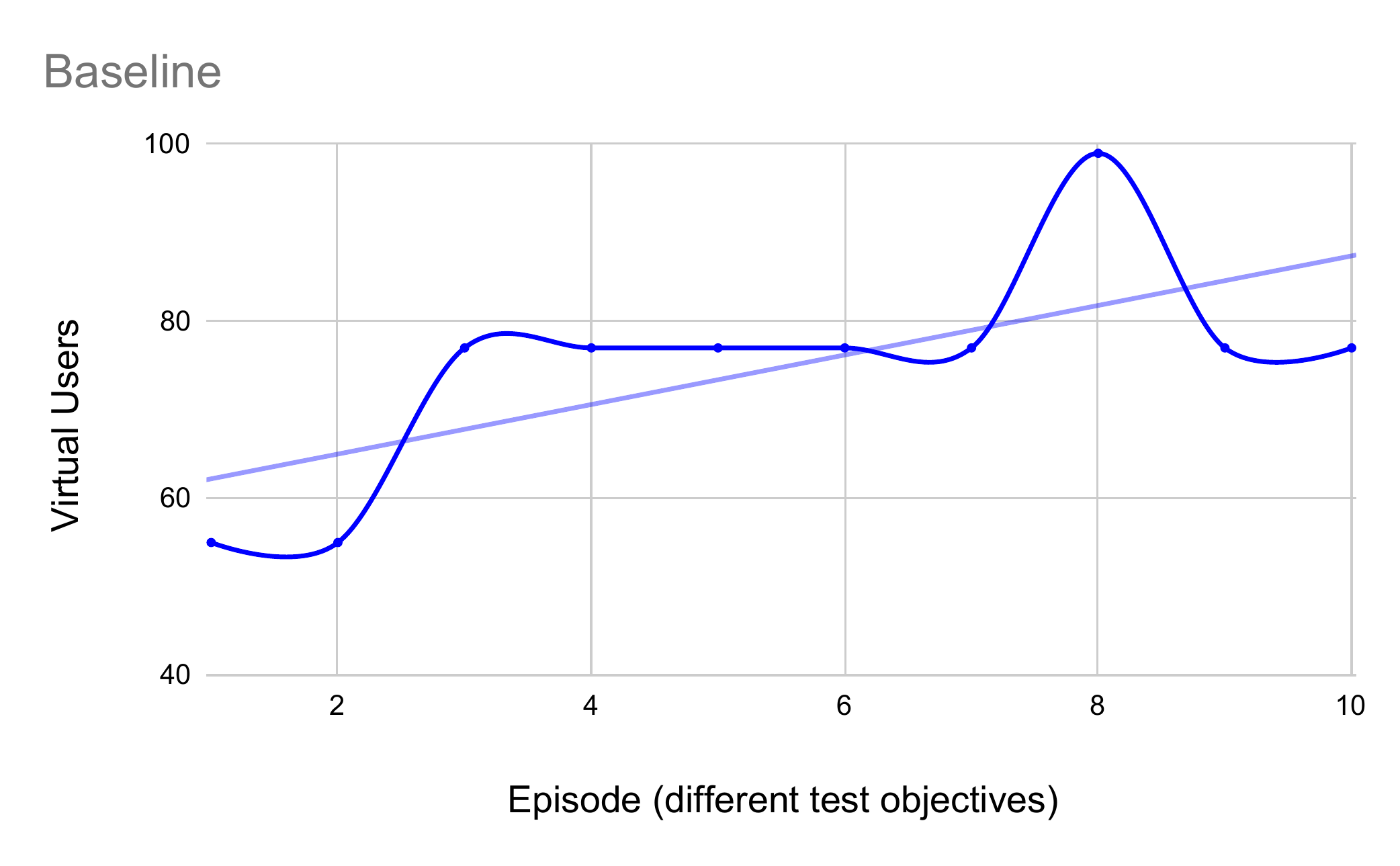}}}
\subfloat[Random Testing]{
    \label{transfer-learning-Random}
    \framebox{\includegraphics[width=0.31\linewidth, height=4cm]{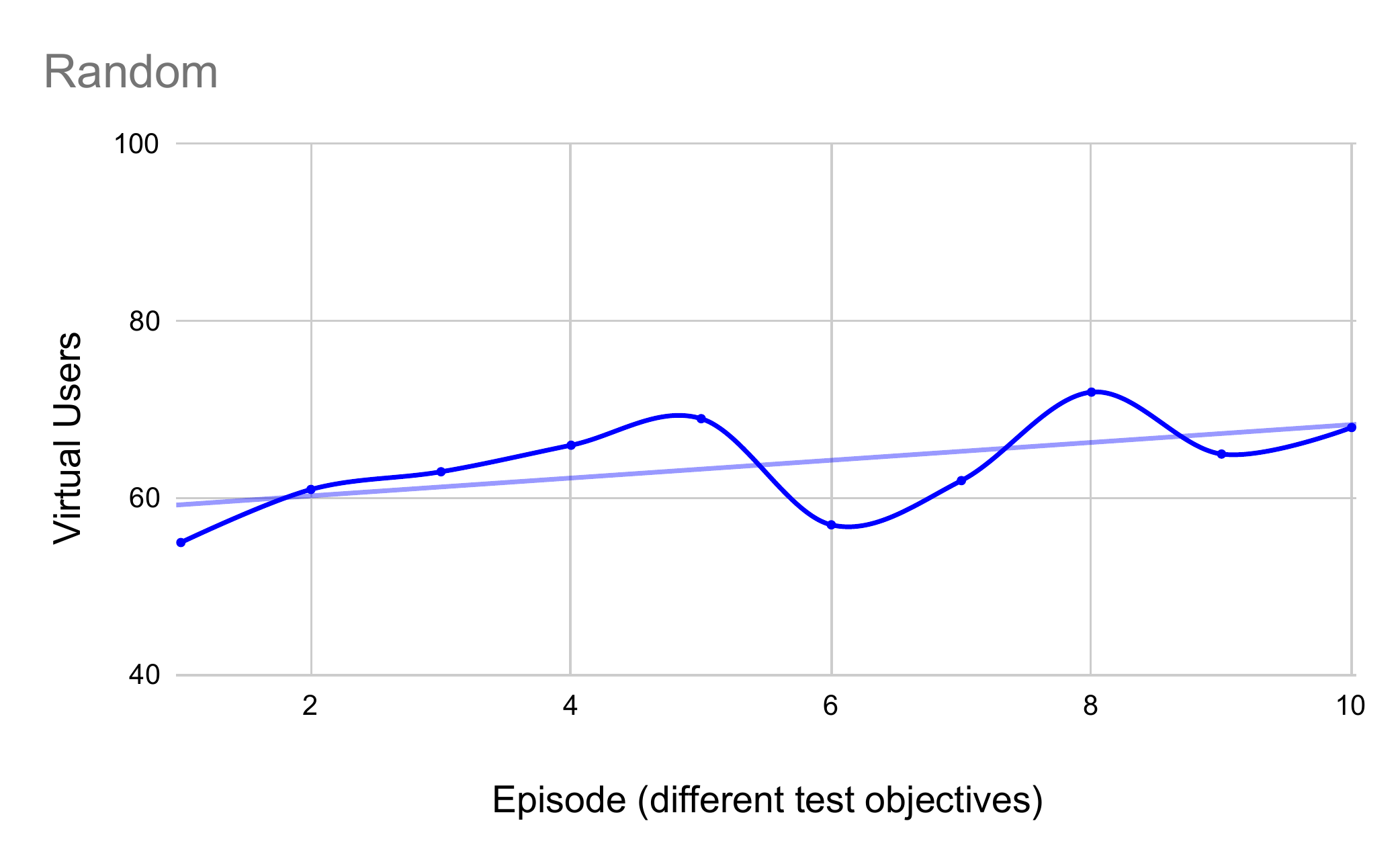}}}
\caption{Efficiency of RELOAD (in the transfer learning) vs. the baseline and random approaches in new similar testing scenarios}
\label{fig: transfer-learning-efficiency}
\end{figure*}

%It provides the highest efficiency over all the learning episodes compared to other variants of the approach with different $\varepsilon$.
\textit{Transfer Learning}. After the initial convergence, we study the efficiency of RELOAD in reusing the learned policy in further similar testing situations (scenarios) during the transfer learning. In this part of the experimentation, after an initial learning of 40 episodes with RELOAD configuration A3, we continue with 10 additional episodes \green{(i.e., episodes 41-50 in Fig. \ref{transfer-learning-Qlearning})}. For these 10 episodes, we change the test objective and keep the $\varepsilon$ low to guide the agent towards relying on the learned policy. Over the episodes of transfer learning, we alter the threshold of the target performance status (i.e., test objective). We change the target error rate threshold from $0.2$ to $0.3$ gradually by an increase of $0.01$ at each episode and also change the target threshold for response time from $1,500 ms$ to $2,500 ms$ by an increase of $100 ms$ at each episode. Figure \ref{fig: transfer-learning-efficiency} shows the efficiency of RELOAD in accomplishing the test objective in the further similar testing scenarios \green{(i.e., represented by the 10 episodes, 41-50, after the initial learning)} compared to the Standard Baseline and Random Testing. It indicates that the smart test agent is able to properly reuse the learned policy in the similar testing scenarios, i.e., the episodes with new values of test objectives, and still accomplish the test objective more efficiently. Table \ref{table: transfer learning-average efficiency} presents the resulting test cost reduction of RELOAD in the transfer learning. 

%It leads to around $\approx 15\% $ test cost saving in the new testing situations (See Table \ref{table: average efficiency}).
%makes an efficiency improvement of around $\approx 13\% $ in the size of the generated workload in the new testing situations. 

%which leads to higher efficiency in generating the proper test workload.
%Despite all the varying conditions on the hosting, it tries to adapt the test workload well to meet the intended objective.
%to be able to hit the error rate threshold without prior knowledge of the SUT system model.

%The baseline approach leads to the worst workload size among all the approaches, while, the random selection of actions in the random load testing  causes a variety of values for the final workload sizes with a constant average (as the trend line in Figure \ref{final_workload_random} shows).
%less required effort for performance assurance of the SUT.
 
%average size of the generated workload and the number of required learning trials or steps for generating the target workload. 

\begin{table}
\centering
\caption{Average test cost saving of RELOAD in the initial learning}
\begin{tabular}{ |m{2.5cm}|m{1.1cm}|m{1.1cm}|m{1.1cm}|m{1.1cm}|}
\hline
\textbf{} & \multicolumn{4}{|c|}{RELOAD} \\
\hline
\textbf{Test Cost Saving} & $\varepsilon=0.5$ & $\varepsilon=0.2$ & \small decaying $\varepsilon$ & DQN setup \\
\hline
\hline
%\textbf{\small Average workload size of the last 10 episodes} & 45 & 45 & 43 &  43\\
%\hline
\textbf{\small w.r.t Standard Baseline} & 30\%  & 30\%  & 34\%  & 34\% \\
\hline
\textbf{\small w.r.t Random Testing} & 17\%  & 17\%  & 20\%  & 20\% \\

\hline
\end{tabular}
\label{table:Initial learning-average efficiency}
\end{table}

%Figure \ref{final_workload_decaying_epsilon} indicates that the agent has learned an optimal policy with regard to the convergence of the workload size to a lower value over the learning episodes.\\
%\textcolor{red}{Figure \ref{fig: learning trials} shows the output of the learning trials in a learning episode (for example using $\varepsilon$-greedy with $\varepsilon = 0.85$) and presents how the RL-assisted approach finds an efficient workload reaching the intended error rate in a number of learning trials.}

\begin{table}
\centering
\caption{Efficiency and average test cost saving in the transfer learning}
\begin{tabular}{ |m{2.5cm}|m{1.9cm}|m{1.4cm}|m{1.4cm}|}
\hline
\textbf{}& \textbf{RELOAD (with Q-learning)}& \textbf{Standard Testing} & \textbf{Random Testing}\\
\hline
%\textbf{Approach} & $\varepsilon=0.5$ & $\varepsilon=0.2$ & \small decaying $\varepsilon$\\
\hline
\textbf{\small Range of the number of generated virtual users} & 48-62 & 55-99 & 55-68\\[5pt] 
\hline
\multicolumn{2}{|c|}{} & & \\
\multicolumn{2}{|c|}{ \textbf{ RELOAD test cost saving}} &  $25\% $ & $ 13\% $\\[5pt]
%\textbf{\small improvement} & 0.15\% & 0.17\% \\ 
\hline
%\textbf{\small Average number of trials of last 10 episodes} & 11.6 & 8.1 & 9.2 & 1.9 & 12.2\\
%\hline
\end{tabular}
\label{table: transfer learning-average efficiency}
\end{table}

\textbf{\textit{Sensitivity Analysis.}} We select $\varepsilon$-greedy with decaying $\varepsilon$ (A3) as the learning configuration in the sensitivity analysis. %We examine the effects of learning rate ($\alpha$) and discount factor ($\gamma$) on the behavioral performance of the approach in a systematic way. 
Figure \ref{fig:sensitivity} shows the behavioral performance of RELOAD regarding changing the values of hyperparameters as described in Section~\ref{sec:method}. It presents how different values for the learning hyperparameters influence the learning behavior, e.g., convergence, and the learning trend, in the proposed RL-driven test agent. We observe that RELOAD does not converge using a low learning rate, i.e., $\alpha=0.1$. Furthermore, we find slower convergence using both lower and higher discount rates, i.e,. $\gamma=0.1$ and $\gamma=0.9$. % In the sensitivity analysis, we consider $\alpha=0.5$ and $\gamma=0.5$ as the baseline values and in four controlled experiments, we study the effects of changing the values of these parameters. First, we set the learning rate to $0.1$ and $decaying$ values while keeping the value of the discount factor fixed at $0.5$, then in the second set of experiments, we set the discount factor to $0.1$ and $0.9$, while fixing the learning rate at $0.5$. 
 
\subsection{Revisiting the Research Questions}
\textbf{\textit{RQ1.}} 
%We studied the efficiency of RELOAD compared to a typical load testing procedure through multiple experiments.
As shown in Fig. \ref{fig:efficiency} and Table \ref{table:Initial learning-average efficiency}, on average RELOAD leads to accomplishing the test objective using fewer virtual users than the Standard Baseline and Random Testing. %After reaching the initial convergence (e.g., during the last 10 episodes)
RELOAD learns how to meet the objective with a more accurate and fine-tuned workload and subsequently leads to a considerable test cost saving. In particular, RELOAD based on $\varepsilon$-greedy, decaying $\varepsilon$ and DQN setup, offers a smoother learning trend and after the convergence results in a slightly higher cost saving, i.e., $34\%$ and $20\%$, compared to other learning configurations. Based on our experiments, we conclude that RELOAD results in 10-30\% increased test efficiency.  

%The results (See Fig. \ref{fig:efficiency} and Table \ref{table:Initial learning-average efficiency}) show that RELOAD during the initial learning learns how to accomplish the test objective efficiently, as after the learning convergence it leads to around $\approx 17\% $ test cost saving.
%presents the resulting efficiency improvement at the last 10 episodes of the proposed approach compared to the baseline approach (i.e., the last 10 episodes show the efficacy of the RL approach when it has almost realized the initial convergence).
The test agent learns the optimal policy to meet the test objective efficiently over the learning episodes. The optimal policy is learned through the value function. The agent stores the learned value function and is able to exploit it in further testing scenarios. The results of the efficiency analysis in the transfer learning (see Fig. \ref{fig: transfer-learning-efficiency} and Table \ref{table: transfer learning-average efficiency}) confirm that the test agent after the initial learning is able to reuse the gained knowledge in subsequent testing scenarios, in which the SUT displays similar characteristics, and maintain its efficiency across scenarios. As shown in Table \ref{table: transfer learning-average efficiency}, RELOAD leads to $25\%$ and $13\%$ test cost saving in the transfer learning compared to the Standard Baseline and Random Testing.  

\textbf{\textit{RQ2.}} As shown in Fig. \ref{fig:sensitivity}, in the sensitivity analysis experiments, fixing the learning rate at a low value such as $0.1$ did not lead to a learning convergence. Whereas a higher value such as $0.5$ (as used in efficiency analysis) or using a decaying learning rate results in faster updates in the stored Q-table of the agent and works better in this case study. %and appears to work better in adaptation to highly changeable environments. 
%It leads to a smooth and partially equal performance during the learning episodes.
Moreover, changing the values of the discount factor, e.g., setting it to $0.1$ or $0.9$,  appears to slow down the learning convergence.

\textbf{\textit{Applicability.}} RELOAD learns how to tune the transactions optimally in the workload to meet the test objective. The smart agent generates an effective workload efficiently without relying on source code or a system model. %and is able to reuse the learned policy in further testing situations. 
%generates a proper workload to meet the intended objective more efficiently than a typical load testing procedure. 
%It generates a more accurate, fine-grained, and smaller workload (in terms of the number of users) in comparison to a typical load testing process, which results in saving more time and cost in the testing.
It is well-suited to operational contexts where the source code, system models, and behavior specifications are not available. Meanwhile, the pay-as-you-go cost for many of the load generation tools on the market is proportional to the number of generated virtual users. Therefore, the efficient generation of an effective workload by the proposed test agent could lead to considerable cost and time savings in the testing process. 
Moreover, the proposed smart test agent has the capability of reusing the learned policy in further similar testing scenarios. %e.g., continuous performance testing scenarios within DevOps. %for example, reaching different target breaking points.
%It reuses the learned policy during the transfer learning. Meanwhile, 
RELOAD keeps the learning running to adapt the learned policy to changes in the environment. This feature is particularly beneficial to DevOps continuous testing activities such as performance regression testing where performance testing scenarios must be repeated for the SUT in a continuous integration process. 

\subsection{Threats to Validity}
Some of the potential sources of threats to validity of the experimental results are described as follows:
%Load scenario generation is heavily dependent on the hardware of the SUT and its execution environment. External factors (such as running the SUT on a shared hosting server) might alter the results. In this section, we address the potential threats to validity based on the classification presented by Runeson and H\"ost~\cite{RuHo09}.
%cite the guidlines for casestudy
%\todoB{ Using a single subject in the experiment e.g. one SUT can be considered as a generalization validity threat.}\\

\textit{Construct validity.} One of the main sources of threat is the formulation of the RL technique to address the problem. Formulating the states, actions, and also the reward function is a major step in building an RL-driven smart agent. %that can learn the optimal policy for meeting an objective properly.

%This aspect of validity reflects to what extent the studied operational measures really represent what the researcher has in mind. A misunderstood question is an example of a potential construct validity threat. To tackle potential threats to our construct validity, we have benefited the guides from multiple researchers in the problem formulation and used well-established guidelines for conducting our study.\\
%to review
\textit{Internal validity.} 
Dependency on the resource availability in the execution environment of the SUT is another common source of threats to the validity of the results in performance testing. To tackle this potential threat, we perform the experiments on dedicated virtual machines, i.e., two separate VMs were used for running the SUT and the test agent. %without any other co-located running applications.% However, there might be still such uncontrolled factors, e.g., the operating system's processes, which could affect the validity of the results.

\textit{External validity.} In our case, the approach has been formulated based on a particular e-commerce web-application as SUT, which supports a certain set of transactions. Therefore, in order to apply the approach to other types of applications, e.g., other web-based systems, the involved transactions of the new system should be extracted and included in the set of actions. 
%and subsequently slight changes in the list of actions are made.     

%This aspect of validity is concerned with to what extent it is possible to generalize the findings, and to what extent the findings are of interest to other people outside the investigated case. In our case, the results are obtained by executing the experiment on one SUT, and we, therefore, do not claim that the results can be generalized to other cases. However, we chose open-source and heavily maintained SUT, so that the results can be generalized to other similar cases (WooCommerce-based e-commerce applications). In addition, our results can also be of interest to performance testing researchers and practitioners working with load testing via JMeter.

\begin{figure*}[h]
\label{fig:sensitivity}
\centering
\subfloat[]{
    \label{final_workload_learning_rate0.1}
    \framebox{\includegraphics[width=0.48\linewidth, height=4cm]{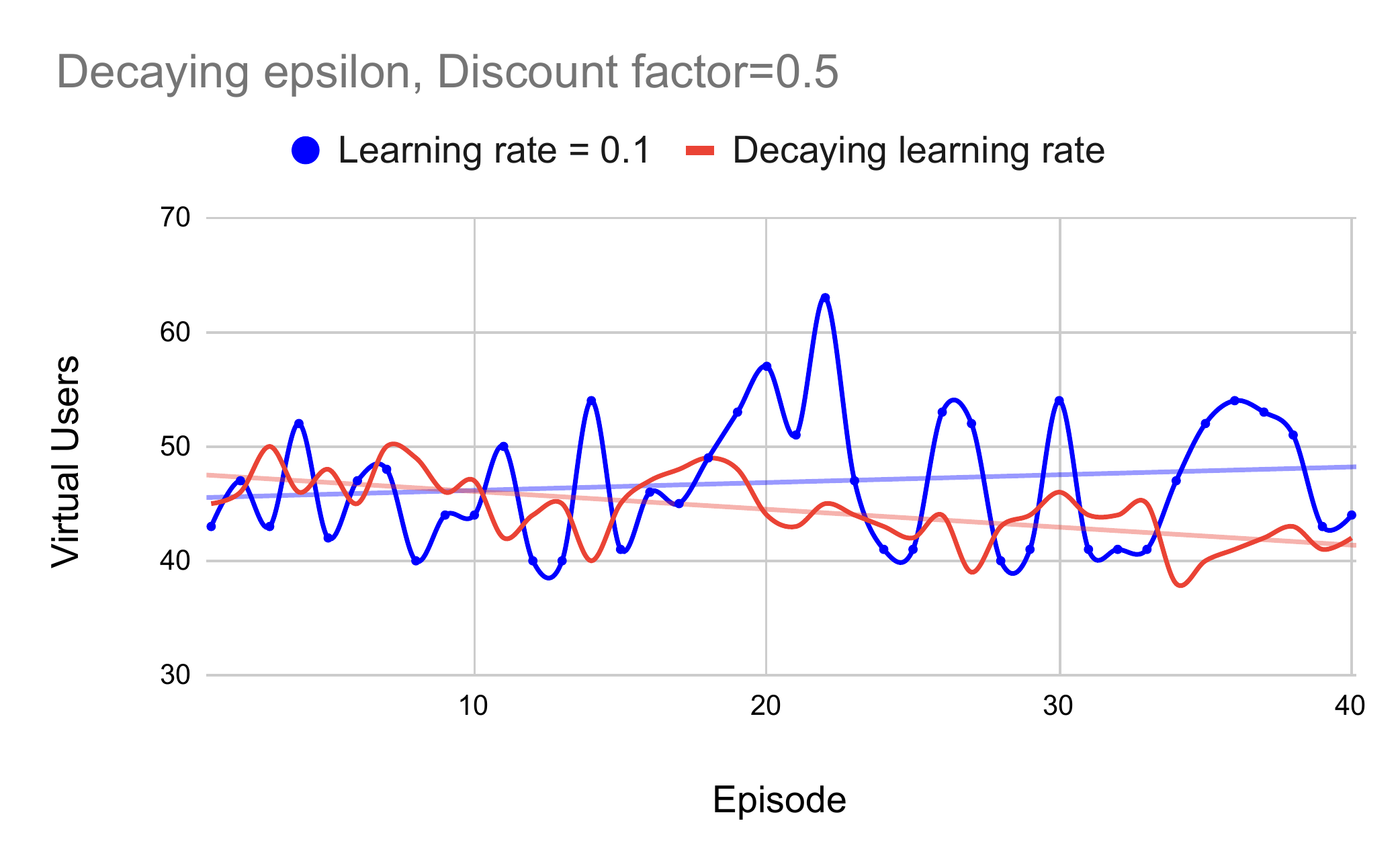}} } 
\begin{comment}
\subfloat[Decaying Learning Rate]{
    \label{final_workload_decaying_learning_rate}
    \framebox{\includegraphics[width=0.48\linewidth, height=4cm]{Figures/_Decaying_epsilon_Decaying_learning_rate_Discount_factor05.pdf}}}
\end{comment}
\subfloat[]{
    \label{final_workload_discount_factor0.1}
    \framebox{\includegraphics[width=0.48\linewidth, height=4cm]{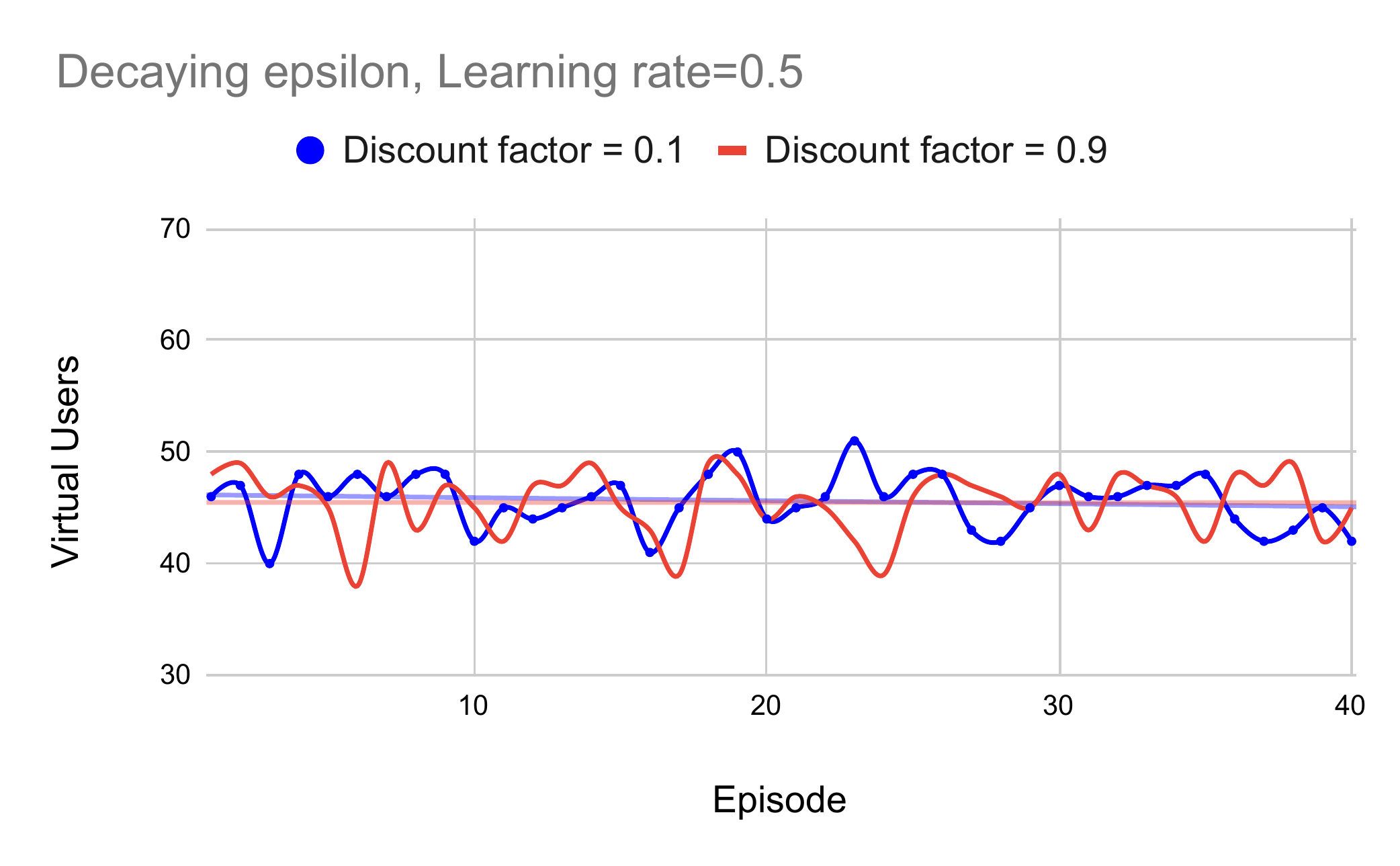}}}
\begin{comment}
\subfloat[Discount Factor = 0.9]{
    \label{final_workload_discount_factor0.9}
    \framebox{\includegraphics[width=0.48\linewidth, height=4cm]{Figures/_Decaying_epsilon_Learning_rate05_Discount_factor09.pdf}}}
\end{comment}
\caption{Behavioral sensitivity of RELOAD to the learning hyperparameters }
\label{fig:sensitivity}
\end{figure*}

\section{Related Work}\label{C:Sec::RealtedWork}
Measuring performance metrics under different execution conditions including various workload and platform configurations \cite{ menasce2002load, apte2017autoperf, jindal2019performance}, detecting different performance-related issues such as functional problems or violations of performance requirements %which emerge under certain workload or resource configuration conditions
\cite{briand2005stress, zhang2011automatic, ayala2018one} are common objectives of different types of performance testing.
%Different approaches to generate the test workload have been proposed in the literature.
An overview of the techniques used for generating test workload is presented as follows:\\  
\textit{Analyzing system models.} Analysis of a performance model of SUT in Petri nets using constraint solving techniques \cite{zhang2002automated}, using genetic algorithms to generate test load based on the control flow graph of SUT \cite{gu2009search}, applying genetic algorithms to other types of system models such as UML models to generate stress test load \cite{garousi2010genetic, garousi2008traffic, costa2012generating, da2011generation} are samples of the techniques in this category.\\
\textit{Analyzing source code.} Generating test load using the analysis of SUT's data-flow and symbolic execution \cite{yang1996towards, zhang2011automatic} are examples of using source code analysis to generate test load and find performance-related issues.\\
\textit{Modeling real usage.} Extracting the usage pattern of real users and modeling their behavior using form-oriented models \cite{draheim2006realistic, lutteroth2008modeling}, extracting workload characteristics and modeling the user behavior based on Extended Finite State Machines \cite{shams2006model} and Markov chains \cite{vogele2018wessbas} through monitoring submitted requests to SUT, and workload characterization through users clustering based on the business-level attributes extracted from usage data \cite{maddodi2018generating} are examples of the techniques used for modeling the realistic workload.\\
\textit{Declarative specification-based methods.} Using a declarative Domain Specific Language (DSL) to specify the performance testing process together with a model-driven test execution framework \cite{ferme2018declarative, walter2016asking} and also using a specific behavior-driven language,  to specify load testing process in combination with a declarative performance testing framework like BenchFlow \cite{schulz2019behavior} are examples of declarative techniques for performance and load testing.\\
\textit{Machine learning-assisted methods.} Machine learning techniques such as supervised and unsupervised algorithms are often intended to build models and knowledge patterns from the data, while in other techniques like reinforcement learning algorithms, the intelligent agent learns the way to accomplish an objective through interaction with the environment. Machine learning techniques have been frequently used for analyzing the resulted data, e.g., for anomaly detection \cite{syer2011identifying} and reliability prediction \cite{avritzer2008reliability}.\\  
%from the load testing, using Bayesian Network to predict the reliability from the load testing data \cite{avritzer2008reliability}, anomaly detection based on analysis of metrics data, e.g., resource usage, using clustering techniques \cite{syer2011identifying}, identifying performance signature based on performance metrics data using supervised and unsupervised learning techniques \cite{malik2013automatic, malik2010automatic} are some examples of using machine learning techniques for analysis of load testing data.
Machine learning techniques have been also applied to the generation of performance test conditions in some studies. For example, using RL together with symbolic execution to find the worst-case execution path within an SUT in \cite{koo2019pyse}, a feedback-driven learning technique which extracts some rules from the execution traces to find the performance bottlenecks, %i.e., the method calls which their execution highly affects the performance
\cite{grechanik2012automatically}, %using RL to find a sequence of input values resulting in performance degradation \cite{ahmad2019exploratory}, 
and using RL to build a smart performance testing framework which mainly generates the platform-based test conditions \cite{moghadam2019machine,moghadam2019machine2,moghadamposter}. %Nonetheless, RL algorithms have been also widely used in performance preservation of software services, such as an adaptive RL-driven performance control for cloud services \cite{ibidunmoye2017adaptive, veni2016auto, jamshidi2016fuzzy} and also software services on other execution platforms \cite{moghadam2018makespan, moghadam2018adaptive}.
Regarding generating performance test conditions, a few studies have also used some other adaptive techniques to generate the test workload. A feedback-based approach using search algorithms to benchmark an NFS server based on changing the test workload in \cite{shivam2008cutting}, and an adaptive generation of test workload based on using some pre-defined tuning policies in \cite{ayala2018one} are some other examples of using adaptive approaches for the generation of performance test conditions.  

\section{Conclusion} \label{C:Sec::Conclusion}
System models, source code, and user behavior patterns are common sources of information in load testing techniques for generating test workload to find performance issues. Nonetheless, those artifacts might not be available all the time during the testing.
Moreover, in black-box testing approaches, it is important to consider that not all transactions have the same effect on the performance, i.e., tuning the workload optimally is crucial for test efficiency. 
We proposed RELOAD, a self-adaptive model-free RL-driven load testing agent that learns how to tune transactions in the workload to accomplish the test objective. It learns an optimal policy to generate an effective workload efficiently and is able to reuse the learned policy in further similar testing scenarios, e.g., in performance regression testing. Furthermore, RELOAD adapts the learned policy to continuous changes in the SUT and the execution environment, thus we believe the smart test agent to be particularly well-suited to the continuous performance testing context within DevOps. The smart test agent assumes two phases of initial and transfer learning and uses Q-learning as the core learning algorithm. It performs more efficiently than random and baseline load testing approaches, which enables reduced testing costs. %such as performance regression testing and continuous testing.  %and is also able to preserve the efficiency in further testing situations and results in time and cost savings which are of great importance to many testing activities such as performance regression testing and continuous testing.   

We conclude that RELOAD provides three main strengths. First, RELOAD provides efficient generation of effective test workloads. Second, the RL approach reduces source code and model dependencies, e.g., system models and user behavior models. Third, RELOAD enables generalizable knowledge representation, i.e., previously learned policies can be reused for other testing scenarios on the SUT. 
%SUTs and execution contexts.
We posit that RELOAD can reduce costs in performance testing. Furthermore,   the continuous testing context that permeates contemporary DevOps processes would further amplify the benefits. In future work, we plan to conduct empirical studies to validate our claims.

%We conclude that the efficient generation of an effective test workload, while reducing dependency on source code, system and user behavior models, the capability of knowledge formation and reusing the knowledge in further testing scenarios, and finally the resulting test cost saving, are the main strengths of the proposed RL-driven load testing agent.
%The value and benefit of the proposed intelligent load testing solution can particularly become more obvious in the context of continuous testing which is prevailing in DevOps practices. %The possibility of testing independently of system and user behavior models and cost reduction is of importance to the testing of complex and evolving software systems.
%of software variants and product lines and also load testing of a SUT on different platforms. 
\balance

%\bibliographystyle{ACM-Reference-Format}
%\bibliography{references} 
\section*{Acknowledgment}
This work has been supported by and received funding from the TESTOMAT and IVVES  European projects.
\bibliographystyle{IEEEtran}
\bibliography{references}

\end{document}